\documentclass{ws-ijmpd}
\usepackage[super,compress]{cite}
\usepackage{graphicx}
\usepackage{epstopdf}
\usepackage{amsfonts}
\usepackage{amssymb}
\usepackage{epsfig}
\usepackage{hyperref}

\usepackage{dcolumn}
\usepackage{bm}
\usepackage{amsmath}
\usepackage{epstopdf}
\usepackage{amsfonts}
\usepackage{amssymb}
\usepackage{epsfig}
\usepackage{tabularx}
\usepackage{color}
\usepackage{hyperref}

\begin{document}


%
\catchline{}{}{}{}{}
%

\title{Charged slowly rotating toroidal black holes in the (1+3)-dimensional Einstein-power-Maxwell theory}

\author{Grigoris Panotopoulos}

\address{Centro de Astrof{\'i}sica e Gravita{\c c}{\~a}o-CENTRA, Departamento de F\'isica, 
\\
Instituto Superior T\'ecnico-IST,
Universidade de Lisboa-UL,
\\
Av. Rovisco Pais 1, 1049-001 Lisboa, Portugal
\\
\href{mailto:grigorios.panotopoulos@tecnico.ulisboa.pt}{\nolinkurl{grigorios.panotopoulos@tecnico.ulisboa.pt}} 
}

\author{{\'A}ngel Rinc{\'o}n}

\address{Instituto de F{\'i}sica, Pontificia Universidad Cat{\'o}lica de Chile,
Av. Vicu{\~n}a Mackenna 4860, Santiago, Chile
\\
\href{mailto:arrincon@uc.cl}{\nolinkurl{arrincon@uc.cl}} }

\maketitle

\begin{history}
\received{Day Month Year}
\revised{Day Month Year}
\end{history}

\begin{abstract}
In this work we find charged slowly rotating solutions in the four-dimensional Einstein-power-Maxwell non-linear electrodynamics assuming a negative cosmological constant. By solving the system of coupled field equations explicitly we obtain an approximate analytical solution in the small rotation limit. The solution obtained is characterized by a flat horizon structure, and it corresponds to a toroidal black hole. The Smarr's formula, the thermodynamics and the invariants Ricci scalar and Kretschmann scalar are briefly discussed.
\end{abstract}


\ccode{PACS numbers: 04.70.Bw; 04.20.-q; 04.70.Dy.}

\section{Introduction}

Within the framework of Einstein's General Relativity (GR) \cite{GR} black holes (BHs) are predicted to exist, and they are very important objects both for classical and quantum gravity. Astrophysical BHs are supposed to be formed during the final stages of massive stars, primordial BHs may be formed in the early Universe from density inhomogeneities, and mini-black holes may be formed at colliders or in the atmosphere of the earth in TeV-scale gravity scenarios in D-brane constructions of the Standard Model \cite{miniBH1,miniBH2,miniBH3}. Furthermore, BHs have temperature and entropy \cite{bekenstein,carter1}, and emit radiation from their horizon \cite{hawking1,hawking2}. Hawking radiation, since it is a manifestation of a quantum effect in curved spacetime, has attracted a lot of interest in the community although it has not been detected in the Universe yet.

The historical LIGO direct detection of gravitational waves \cite{ligo1,ligo2,ligo3} has opened a new window to the Universe, and has offered us the strongest evidence so far that BHs do exist in Nature and they merge. In the light of this development the so-called quasinormal modes $\omega = \omega_R + i \omega_I$, with a non-vanishing imaginary part, have become very relevant, since they do not depend on the initial conditions, and they thus carry unique information on the few BH parameters \cite{Kokkotas,Cardoso,Konoplya}. Therefore it becomes clear that BHs are exciting objects that link together several areas of research, from gravity, to astrophysics, to quantum physics, to thermodynamics and statistical physics, and they have become an excellent laboratory to study and understand various aspects of classical and quantum gravity. 

Soon after the formulation of GR, the Schwarzschild \cite{SBH} as well as the Reissner-Nordstr{\"o}m \cite{RN} solutions were obtained, characterized by the mass and the electric charge, respectively, and much later the Kerr solution \cite{kerr} for rotating BHs was found. The theoretical paradigm of the no-hair theorem \cite{TMW}, according to which BHs are uniquely characterized by its mass, electric and/or magnetic charge and angular momentum (although counter examples do exist. See e.g. the recent reviews \cite{counter1,counter2} on hairy BHs), the Smarr's formula \cite{smarr1,smarr2}, and the area law of the BH entropy \cite{bekenstein,carter1}, are valid in GR coupled to Maxwell's linear electrodynamics in (1+3) dimensions. There are, however, good theoretical reasons to consider alternative theories of gravity, or non-linear electrodynamics or extra dimensions. In all of these cases some of the previous conditions are modified or violated.

Kaluza-Klein theories \cite{kaluza,klein}, supergravity \cite{nilles} and Superstring/M Theory \cite{ST1,ST2} have pushed forward the idea that extra spatial dimensions may exist. In more than four dimensions higher order curvature terms are natural in Lovelock theory \cite{Lovelock}, and also higher order curvature corrections appear in the low-energy effective equations of Superstring Theory \cite{ramgoolam}. Furthermore, gravity in (1+2) dimensions is special and has attracted a lot of attention due to the deep connection to a Yang-Mills theory with the Chern-Simons term only \cite{CS,Witten1,Witten2}, and also due to its mathematical simplicity as there are no propagating degrees of freedom.
What is more, the current cosmic acceleration \cite{SN1,SN2} and the cosmological constant problem \cite{weinberg} have forced us to explore other possibilities modifying the gravitational theory, and studying e.g. $f(R)$ theories of gravity \cite{mod1,mod2} or scalar-tensor theories of gravity \cite{BD} or brane models \cite{dgp}. In addition, non-linear electrodynamics has attracted a lot of attention for several different reasons. Originally the Born-Infeld non-linear electrodynamics was introduced in the 30's in order to obtain a finite self-energy of point-like charges \cite{BI}. During  the  last  decades  this type of action reappears in the open sector of superstring theory as it describes the dynamics of D-branes \cite{Dbranes1,Dbranes2}. On the other hand, straightforward generalization of Maxwell’s theory leads to the so called Einstein-power-Maxwell (EpM) theory described by a Lagrangian density of the form $\mathcal{L}(F) = F^s$, where $F$ is the Maxwell invariant, and $s$ is an arbitrary rational number. Clearly the special value $s = 1$ corresponds to linear electrodynamics. 

Currently, this class of non-linear electrodynamics has been receiving attention in several contexts \cite{Xu:2014uka,Mazharimousavi:2011nd,BH1,BH2,mehrab,Rincon:2018dsq}. The reason why studying such a class of theories is interesting lies on the fact that Maxwell's theory in higher dimensions is not conformally invariant. In \cite{mehrab} it was shown that in higher-dimensional Conformal Gravity in order to obtain charged BH solutions one has to consider the EpM theory, and there is no other option. In a D-dimensional spacetime the electromagnetic stress-energy tensor is traceless if the power $s$ is chosen to be $s=D/4$. Therefore in four dimensions the linear theory is conformally invariant, and this corresponds of course to the standard Maxwell's theory. In a three-dimensional spacetime, however, if $s=1$ the theory is linear but the electromagnetic stress-energy tensor is not traceless, whereas if $s=3/4$ the theory is conformally invariant but non-linear. 

Smarr's formula in non-linear electrodynamics has been studied in \cite{Breton,Fernando}, and in 5D supergravity in \cite{Townsend}. Many black hole solutions, rotating and non-rotating, already exist in the literature, and have been obtained in several different contexts. We mention here the following cases: BH solutions in brane-world models have been obtained in \cite{Kofinas1,Kofinas2}, for a review on BHs in higher curvature gravity see e.g. \cite{Charmousis}, and in $f(R)$ theories of gravity in \cite{Multamaki:2006zb}. The charged rotating BH in (1+2) dimensions has been obtained in \cite{zanelli}, charged rotating BHs in supergravity have been obtained in \cite{cvetic1,cvetic2}, and the higher-dimensional rotating BH has been obtained in \cite{MP}. Finally, charged BH solutions in (1+2)-dimensional and higher-dimensional EpM theories have been obtained in \cite{BH1} and \cite{BH2} respectively.

Slowly rotating BH solutions in non-linear electrodynamics in four dimensions have been obtained in \cite{iran1}, and in EpM theory in particular in any number of dimensions $D \geq 4$ in \cite{iran2}. In all of these cases the solutions are characterized by spherical horizon structure. It is possible, however, to construct rotating solutions with flat horizon structure \cite{lemos1,lemos2}. Our goal here is to obtain a charged rotating solution with flat horizon structure in non-linear electrodynamics. Therefore in the present article we obtain in an analytical way another class of charged slowly rotating toroidal black holes in EpM theory with a negative cosmological constant, generalizing the charged solution studied in \cite{lemos2} in the framework of the Einstein-Maxwell theory.

Our work is organized as follows:  After this introduction, we present the theory and the set of coupled field equations in section \ref{action}, while in the third section we obtain non-rotating charged solutions with flat horizon structure. The slowly rotating charged black hole with toroidal topology is obtained in an analytical way by solving the field equations explicitly in section \ref{Sol}. Finally, we conclude our work in Section \ref{Concl}.

\section{The action and the set of fields equations}
\label{action}

Let us consider the theory in (1+3) dimensions described by the action
\begin{equation}
S[g_{\mu \nu}, A_{\mu}] = \int \mathrm{d} ^4x \sqrt{-g} \left[ \frac{1}{2 \kappa} (R-2\Lambda) - \alpha (F_{\mu \nu} F^{\mu \nu})^s \right]
\end{equation}
where $\kappa=8 \pi G$, with $G$ being Newton's constant, is the gravitational constant, $s$ is an arbitrary rational number, $R$ is the Ricci scalar, $g$ the determinant of the metric, $F_{\mu \nu}$ is the electromagnetic field strength, and $\Lambda$ is the cosmological constant, which for the time being can be either positive or negative. Besides, we recover the linear Maxwell's theory for $s=1$. Varying the action with respect to the potential $A_\mu$ one obtains the generalized Maxwell's equations \cite{BH2}
\begin{equation}
\partial_\mu (\sqrt{-g} F^{\mu \nu} F^{s-1}) = 0
\end{equation}
where $F \equiv F_{\mu \nu} F^{\mu \nu}$ is the Maxwell invariant, while varying with respect to the metric tensor $g_{\mu \nu}$ one obtains Einstein's field equations \cite{BH2}
\begin{equation}
R_{\mu \nu} - \frac{1}{2} R g_{\mu \nu} + \Lambda g_{\mu \nu} = \kappa T_{\mu \nu}^{\text{EM}}
\end{equation}
where $R_{\mu \nu}$ is the Ricci tensor, and $T_{\mu \nu}^{\text{EM}}$ is the electromagnetic tensor
\begin{align}
T_{\mu \nu}^{\text{EM}} &= 4 \alpha \left [s F_{\mu \rho} F_\nu ^\rho F^{s-1} - \frac{1}{4} g_{\mu \nu} F^s \right ]
\\
& = A g_{\mu \nu}+D F_{\mu \rho} F_\nu ^\rho
\end{align}
where we have introduced the quantities
\begin{eqnarray}
A & = & - \kappa \alpha F^s \\
D & = & 4 \kappa \alpha s F^{s-1}
\end{eqnarray}
Taking the trace of Einstein's equations we can express the Ricci scalar in terms of the trace of the electromagnetic stress-energy tensor, and after that the field equations take the form
\begin{equation}
R_{\mu \nu} -\Lambda g_{\mu \nu} = \Theta_{\mu \nu} \equiv (2s-1) g_{\mu \nu} + D F_{\mu \rho} F_\nu ^\rho
\end{equation}

\section{Non-rotating charged solutions}

\subsection{Solutions with spherical horizon structure}

For static solutions without rotation we employ the coordinate system $t,r,\theta,\phi$, we identify the coordinates as $x^{\mu} = \{t,r,\theta,\phi\}$
and we make for the Maxwell's potential the ansatz $A_i=0, A_t(r)$, while for the metric tensor we make the ansatz
\begin{equation}
ds^2 = -f(r) dt^2 + f(r)^{-1} dr^2 + r^2 \gamma_{ij} dx^i dx^j
\end{equation}
where the indices $i,j$ run from 1 to 2, and $\gamma_{ij}$ is the metric of a 2-dimensional surface with constant curvature $k=-1,0,1$. 
In the case where $k=1$ (spherical horizon structure), $\gamma_{ij} dx^i dx^j=d \theta^2 + \sin^2(\theta) d\phi^2$, the solution is given by
\begin{eqnarray}
f_{k=1}(r) & = & 1-\frac{2 M}{r}+\frac{B}{r^\beta}-\frac{\Lambda r^2}{3} \\
F_{tr} & = & \frac{C}{r^\beta}
\end{eqnarray}
where setting the charge of the black hole to zero, $B=0$, we recover the Schwarzschild-(anti) de Sitter solution, while setting $\Lambda=0$ we recover the BH solution obtained in \cite{BH2}
with $M$ being the mass of the black hole, $C$ being a constant of integration, and $\beta,B$ are computed to be
\begin{eqnarray}
\beta & = & \frac{2}{2s-1} \\
B & = & - \kappa \alpha (-2 C^2)^{s} \frac{(2s-1)^2}{3-2s}
\end{eqnarray}
assuming $\beta > 1$ so that the standard Reissner-Nordstr{\"o}m-(anti) de Sitter BH of Maxwell's linear theory, $s=1$, is included as a special case. 
The choice $\alpha=(2 \kappa (-1)^{s+1})^{-1}$ limits to the familiar Gaussian units used in RN, for which $C=Q, B=Q^2$, with $Q$ being the electric charge of the black hole. The electric charge for a generic electromagnetic Lagrangian $\mathcal{L}$ is computed by \cite{Fernando}
\begin{equation}
\left( \frac{\partial \mathcal{L}}{\partial \bar{F}} \right) E = \frac{Q}{r^2}
\end{equation}
with $E=F_{tr}$ being the electric field, and $\bar{F}=F/4$. We find 
\begin{eqnarray}
C & = & \left( \frac{Q}{s 2^{s-1}} \right)^{\frac{1}{2s-1}} \\
B & = & 2^{s-1} \frac{(2s-1)^2}{3-2s} \left( \frac{Q}{s 2^{s-1}} \right)^{\frac{2s}{2s-1}}
\end{eqnarray}
which for $s=1$ are reduced to $C=Q$ and $B=Q^2$.

The thermodynamics of this class of BHs has been studied in \cite{thermoBH}, while the Smarr's formula for this non-linear electrodynamics takes the form \cite{Breton,Fernando}
\begin{equation}
M = 2 T_H S + \frac{1}{s} Q \Phi_H
\end{equation}
where $S$ is the entropy, $T_H$ is the Hawking temperature, and $\Phi_H$ is the electrostatic potential evaluated at the horizon. Clearly, for $s=1$ we recover the standard formula for the Reissner-Nordstr{\"o}m-(anti) de Sitter BH \cite{smarr2}.

\subsection{Solutions with flat horizon structure}

In the present work we shall consider the case where $k=0$ and in which $\gamma_{ij} dx^i dx^j=d \theta^2 + d\phi^2$. In this case it is advantageous to slightly change the notation, and adopt a coordinate system $t,r,\phi,z$ with a line element for the metric of the form \cite{lemos3}
\begin{equation}
ds^2 = -f(r) dt^2 + f(r)^{-1} dr^2 + r^2 d \phi^2 + \frac{r^2}{l^2} dz^2
\end{equation}
where the range of the coordinates $\phi,z$ determines the kind of the black hole (cylindrical, planar or toroidal) \cite{lemos3}.
Then the diagonal components of the Ricci tensor are computed to be
\begin{eqnarray}
R_{tt} & = & \frac{f(r)}{2} \left( f''(r) + \frac{2f'(r)}{r} \right) \\
R_{rr} & = & - \frac{r f''(r) + 2 f'(r)}{2 r f(r)}  \\
R_{\phi \phi} & = & -f(r)-r f'(r)
\end{eqnarray}
and $R_{zz}=R_{\phi \phi}$, where the prime denoted differentiation with respect to the radial coordinate. Furthermore, the diagonal components of the tensor $\Theta_{\mu \nu}$ are given by
\begin{eqnarray}
\Theta_{tt} & = & (4s-1) \alpha \kappa f(r) (-2)^s (q')^{2s} \\
\Theta_{rr} & = & - \frac{(4s-1) \alpha \kappa (-2)^s (q')^{2s}}{f(r)}  \\
\Theta_{\phi \phi} & = & - (2s-1) \alpha \kappa r^2 (-2)^s (q')^{2s}
\end{eqnarray}
and $\Theta_{zz}=\Theta_{\phi \phi}$. 

The generalized Maxwell's equation gives us for $F_{tr}$ the same solution as before in the $k=1$ case, while using the $R_{\phi \phi}=\Theta_{\phi \phi}$ field equation, the metric function is now computed to be
\begin{eqnarray}
f_{k=0}(r) & = & -\frac{\mu}{r}+\frac{B}{r^\beta} - \frac{\Lambda r^2}{3}\\
F_{tr} & = & \frac{C}{r^\beta}
\end{eqnarray}
and $\beta, B$ are still given by the previous expressions, while the mass scale $\mu$ is related to the conserved mass $M$ of the black hole.

When the charge is zero, $B=0$, a horizon exists only for a negative cosmological constant, $\Lambda = -3/l^2$. Therefore in the following we shall consider this case. When the black hole is charged there are a Cauchy (inner) and an event horizon, $r_-, r_H$ respectively. The solution presented here is the generalization of the solution considered in \cite{lemos2,thermoString} in the framework of the Einstein-Maxwell theory. Its properties were discussed in detail, and the Penrose-Carter diagrams \cite{carter2,ellis} were drawn as well in \cite{lemos2}, while its thermodynamics was studied in \cite{thermoString}.

When $z$ is compact, $0 \leq z < 2 \pi l$, both $\phi$ and $z/l$ take values in the range $[0, 2 \pi)$, and the solution considered here corresponds to a toroidal black hole \cite{lemos3,extra} with surface area $A_H = 4 \pi^2 r_H^2$ \cite{torus}.
Furthermore, the mass parameter $\mu$ and the conserved mass $M$ are related via $\mu=2 M/\pi$ \cite{torus}. The Bekenstein-Hawking entropy $S$ still satisfies the area law
\begin{equation}
S = \frac{1}{4}\mathcal{A}_H = \pi^2 r_H^2
\end{equation}
What is more, the Hawking temperature is given by
\begin{equation}
T_H = \frac{1}{4 \pi} \: f_{k=0}'(r_H) 
\end{equation}
where the event horizon $r_H$ is determined solving the algebraic equation $f_{k=0}(r_H)=0$, which implies
\begin{equation}
\frac{r_H^3}{l^2} = \mu - \frac{B}{r_H^{\beta-1}}
\end{equation}
and then the Hawking temperature takes the form
\begin{equation}
T_H = \frac{1}{4 \pi r_H^2} \: \left( 3 \mu - \frac{\beta + 2}{r_H^{\beta-1}} \: B \right)
\end{equation}
Combining the expressions for $S$ and $T_H$ we obtain the new Smarr's formula for a charged toroidal black hole, which reads
\begin{equation}
\boxed{3 \mu = \frac{4}{\pi} T_H \: S + \frac{B}{C} \: (\beta-1) (\beta + 2) \: \Phi_H}
\end{equation}
which is our first main result in this work, and which for the linear theory $s=1$ takes the simple form
\begin{equation}
3 M = 2 T_H \: S + 2 \pi \: Q \: \Phi_H
\end{equation}
We see that it has the usual structure "mass term = thermodynamics + charge term", although now the numerical prefactors are different compared to the $k=1$ case.

\section{Charged slowly rotating toroidal black holes in EpM }
\label{Sol}

To find stationary axisymmetric solutions we make for the Maxwell's potential the ansatz
\begin{eqnarray}
A_t & = & q(r) \\
A_r & = & 0 \\
A_\phi & = &  a \: p(r) \\
A_z & = & 0
\end{eqnarray}
while for the line element we make the ansatz
\begin{equation}\label{metric_full}
ds^2 = -f(r) dt^2 + f(r)^{-1} dr^2 + r^2 d \phi^2 + \frac{r^2}{l^2} dz^2 + 2 a h(r) dt d\phi
\end{equation}
where $a$ is the rotation parameter assumed to be small. We shall be working in leading order in $a$ neglecting all terms of second and higher order.

It is straightforward to compute the components of the Ricci tensor, of the electromagnetic field strength tensor, and of $\Theta_{\mu \nu}$ at leading order in $a$. We find that the non-vanishing components of $F^{\mu \nu}$
are the following
\begin{eqnarray}
F^{rt} & = & q'(r) \\
F^{\phi r} & = & a \: \frac{h q' + f p'}{r^2}
\end{eqnarray}

The non-vanishing components of $\Theta_{\mu \nu}$ and $R_{\mu \nu}$ are as follows: The diagonal components are the same as when $a=0$ plus corrections of second order which are neglected, and we thus obtain for $f(r)$ the same metric function as in the $a=0$ case
\begin{equation}\label{f_relevante}
f(r) = f_{k=0}(r) = -\frac{\mu}{r}+\frac{B}{r^\beta} + \frac{r^2}{l^2}
\end{equation} 
The only non-diagonal component is the $t \phi$ one, which is computed to be
\begin{eqnarray}
R_{t \phi} & = & - \frac{a}{2} \: \left( f \: h'' + \frac{2 \: h \: f'}{r} \right) \\
\Theta_{t \phi} & = & a \frac{\alpha \kappa (-2)^s (q')^{2s} [2 s f p' - (2s-1) h q']}{q'}
\end{eqnarray}
Regarding the generalized Maxwell's equations, two of them are automatically satisfied, while out of the two non-trivial ones the first equation reads
\begin{equation}\label{eqdif_q}
2q'(r)+(2s - 1) r q''(r) = 0
\end{equation} 
which can be easily integrated to obtain
\begin{equation}
q(r) = C_1 + \frac{C}{(\beta-1) r^{\beta-1}}
\end{equation}
with two integration constants $C, C_1$. However, demanding that
\begin{align}
\lim_{r \rightarrow \infty} q(r) &= 0
\end{align}
$C_1$ is taken to be zero, and we thus obtain the $F^{tr}=C/r^\beta$ of the non-rotating case.
Therefore in total we are left with the following two equations for the unknown functions $p(r), h(r)$
\begin{eqnarray}
R_{t \phi} & = &  \Theta_{t \phi} +\Lambda g_{t \phi} \\
0 & = & \frac{\partial }{\partial r} (r^2 F^{s-1} F^{r \phi})
\end{eqnarray}
The equations above are satisfied taking $p(r)=-q(r)$ and $h(r)=f_{k=0}(r)-r^2/l^2$. This concludes the construction of the solution, and we thus find that the unknown functions are computed to be
\begin{equation}
\boxed{A_{\phi}  = - a \: \frac{C}{(\beta-1) r^{\beta-1}}}
\end{equation}
\begin{equation}
\boxed{h(r) = -\frac{\mu}{r} + \frac{B}{r^\beta}}
\end{equation}
which is our second main result in this work. 

It is easy to verify that the black hole solution presented and discussed in \cite{lemos2,extra}, when all quadratic terms in $a$ are neglected, is reduced to the solution obtained here for $s=1$. Note, however, that in \cite{lemos2,extra}, since the authors in these works discuss a black string with cylindrical topology, $M,Q$ are the mass and the electric charge, respectively, per unit length along the cylinder axis.

We can investigate the effect of rotation on the invariants as well. In particular we will focus on two of them, i.e the Ricci scalar $R$ and the Kretschmann scalar $\mathcal{K}$. Given the metric \eqref{metric_full}, we obtain at leading order in $a$
\begin{align}
R &= -f''(r)-\frac{2 \left[2 r f'(r)+f(r)\right]}{r^2} + \mathcal{O}(a^2)
\label{R_full}
\\
\mathcal{K} &=f''(r)^2+\frac{4 \left[r^2 f'(r)^2+f(r)^2\right]}{r^4}+ \mathcal{O}(a^2)
\label{K_full}
\end{align}
We see that the parameter $a$ is absent at first order. After the replacement of Eq. \eqref{f_relevante} into Eqs. \eqref{R_full} and \eqref{K_full} we explicitly have
\begin{align}
R = \ &4 \Lambda -(\beta -2) (\beta -1) \frac{B}{r^{\beta +2}}
\\
\begin{split}
\mathcal{K} = \ &\frac{1}{3} r^{-2 (\beta +3)} 
\Bigg[
3 \left((\beta  (\beta +2)+5) \beta ^2+4\right) B^2 r^2
-
4 B r^{\beta +1}  \times
\\
& \left(3 (\beta +1) (\beta +2) \mu +(\beta -2) (\beta -1) \Lambda  r^3\right)
+
4 r^{2 \beta } \left(9 \mu ^2+2 \Lambda ^2 r^6\right)
\Bigg]
\end{split}
\end{align}
and we observe that no singularities appear due to the slow rotation.

Before we finish a few remarks are in order. Since the metric function $f_{k=0}(r)$ remains the same as in the non-rotating case, both the entropy $S$ and the Hawking temperature $T_H$ are still given by the same formulas. What is more, the presence of the angular momentum in general modifies the Smarr's formula, in which an extra term is added \cite{smarr2}. The extra term, however, is of second order in the rotation parameter $a$, whereas in this article we work in leading order. Therefore, the extra term due to the non-vanishing angular momentum is negligible, and the Smarr's formula remains intact in the case of the slowly rotating BH solutions, which can be easily verified explicitly.
Finally, it would be interesting to investigate how the properties of the solution obtained here are modified in the framework of the so called scale-dependent scenario, where the coupling constants acquire a dependence on the scale, i.e.  $\{G_0,\Lambda_0\} \rightarrow \{G_k, \Lambda_k\}$), and which has received considerable attention lately
\cite{Koch:2016uso,Rincon:2017ypd,Rincon:2017goj,Rincon:2017ayr,
Contreras:2017eza,Hernandez-Arboleda:2018qdo,Contreras:2018dhs,Rincon:2018lyd,Rincon:2018sgd,Contreras:2018swc}. We hope to be able to address that issue in a future work.

\section{Conclusions}
\label{Concl}

We have constructed charged slowly rotating solutions with flat horizon structure, corresponding to toroidal black holes, in the four-dimensional Einstein-power-Maxwell non-linear electrodynamics including a negative cosmological constant. Under the only approximation that the rotating parameter $a$ is small, we have explicitly computed the non-vanishing components of the Ricci tensor in leading order in $a$, and we have solved analytically the system of coupled equations for the metric functions as well as for the Maxwell's potential. The thermodynamics, the Smarr's formula as well as the two invariants, Ricci scalar $R$ and Kretschmann scalar $\mathcal{K}$, are briefly discussed.


\section*{Acknowlegements}

The authors wish to thank Jos{\'e} P. S. Lemos for disussions as well as the anonymous reviewer for suggestions to improve the presentation of the manuscript.
G. P. thanks the Funda\c c\~ao para a Ci\^encia e Tecnologia (FCT), Portugal, for the financial support to the Center for Astrophysics and Gravitation-CENTRA,  Instituto Superior T\'ecnico,  Universidade de Lisboa,  through the Grant No. UID/FIS/00099/2013. 
The author A. R. was supported by the CONICYT-PCHA/\- Doctorado Nacional/2015-21151658.


\end{document}